# Optoelectronically probing the density of nanowire surface trap states to the single state limit

Yaping Dan

University of Michigan – Shanghai Jiao Tong University Joint Institute, Shanghai, China 200240

Correspondence should be addressed to: yaping.dan@sjtu.edu.cn

**Due to the large surface-to-volume ratio, surface trap states play a dominant role in the optoelectronic properties of nanoscale devices(*1-6*). Understanding the surface trap states allows us to properly engineer the device surfaces for better performance. But characterization of surface trap states at nanoscale has been a formidable challenge using the traditional capacitive techniques based on metal-insulator-semiconductor (MIS) structures(*7*) and deep level transient spectroscopy (DLTS)(*8-11*). Here, we demonstrate a simple but powerful optoelectronic method to probe the density of nanowire surface trap states to the limit of a single trap state. Unlike traditional capacitive techniques (Fig1a), in this method we choose to tune the quasi-Fermi level across the bandgap of a silicon nanowire photoconductor, allowing for capture and emission of photogenerated charge carriers by surface trap states (Fig1b). The experimental data show that the energy density of nanowire surface trap states is in a range from $10^9$ cm$^{-2}$/eV at deep levels to $10^{12}$ cm$^{-2}$/eV in the middle of the upper half bandgap. This optoelectronic method allows us to conveniently probe trap states of ultra-scaled nano/quantum devices at extremely high precision.**

To understand this method, let us use p-type semiconductors as an example. A highly doped p-type semiconductor has an extremely high concentration of holes but a very low concentration of electrons. Under illumination, excess electrons and holes are generated in the conduction and valence band, respectively. At small injection condition, these excess carriers are negligibly low in concentration compared to the majority holes, but often orders of magnitude larger than the minority electrons. Consequently, the quasi-Fermi level of electrons ($E_F^n$) shifts away significantly from the Fermi level at equilibrium ($E_F$), while the quasi-Fermi level of holes ($E_F^p$) remains nearly unchanged as $E_F$, as shown in Fig.1b. In general, the shift of quasi-Fermi levels leads to filling the trap states below $E_F^n$ (but above $E_F$) with electrons, and those above $E_F^p$ (but below $E_F$) with holes. Clearly, for a highly doped p-type semiconductor at small injection condition, only the minority electrons get involved in the capture-emission process of trap states. For every photogenerated electron captured by trap states, there is one corresponding hole remaining in the valence band to contribute to photoconductance. The shift of $E_F^n$ will allow a large number of electrons to be trapped if the density of trap states, at deep levels in particular, is high, leading to the giant gain in photoconductance that has been widely observed(*1-3*).

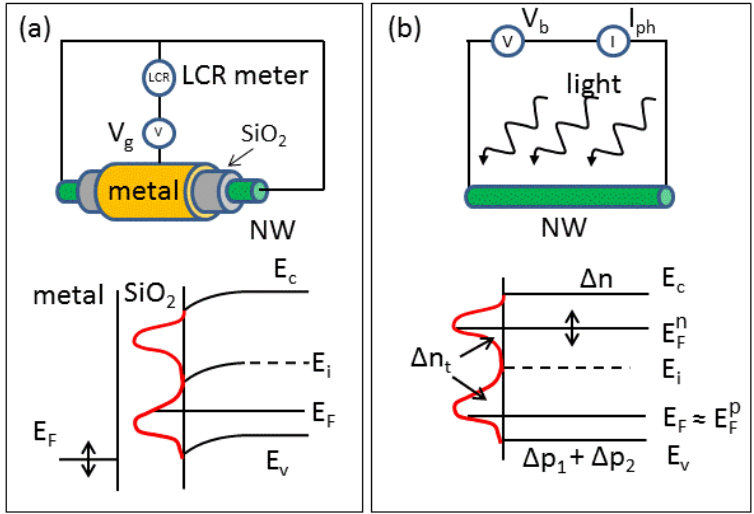

Figure 1. Methods for probing the density of surface trap states. (a) Metal-Insulator-Semiconductor (MIS) capacitor. The gate voltage drives the Fermi level to sweep across the bandgap on the Si/SiO$_2$ interface to allow for capture and emission of charges carriers by surface trap states. (b) Quasi-Fermi level can be tuned across the bandgap by changing the intensity of optical illumination. Δn is the excess electron concentration in the conduction band, and Δn$_t$ the excess electron concentration captured by the surface trap states. The excess carriers in the valence band can be divided into two parts. One part (Δp$_1$) corresponds to the trapped electrons Δn$_t$ and the other (Δp$_2$) is equal to Δn. We have Δn$_t$ + Δn = Δp$_1$ + Δp$_2$.

When the illumination is turned off, the excess electrons in the conduction band and those captured by the trap states will all recombine with the excess holes in the valence band eventually, but the recombination processes are different. For the electrons in the conduction band, the recombination occurs almost immediately through surface recombination centers (~ 100 picoseconds for unpassivated nanowires(*12, 13*)). In contrast, the electrons in the trap states are first emitted to the conduction band at a much lower rate (milliseconds to tens of seconds) and then recombine with the corresponding holes. With this contrast in time domain, we can decouple these two processes and separately extract photogenerated electrons in the conduction band Δn and those in the trap states Δn$_t$ by tuning the modulation frequency of illumination. At high frequencies, for instance, Δn$_t$ will not be able to keep up the pace of the modulation, leaving only Δn to contribute to the photoconductance. The location of $E_F^n$ can be found from Δn. Differentiating Δn$_t$ with respect to $E_F^n$ will allow us to obtain information about the energy density of trap states.

Following the argument above, we first conduct four-probe measurements on a single nanowire device (L=2.4 μm long and 80 nm in diameter) that is Ohmically contacted by four microelectrodes (Fig.2a). The nanowire is p-type and the doping concentration is ~10$^{18}$ cm$^{-3}$. Then we place the nanowire device under illumination of a monochromatic green light beam (λ

~ 500nm) that is modulated on/off by a mechanical chopper. A lock-in amplifier is employed to pick up the periodic ac photocurrent. The transient behavior of the nanowire conductance under the periodic on/off illumination can be described by the red curve in Fig.2b. During the first half period which is under illumination, the conductance starts with an instantaneous jump by $\sigma_{ph}$ followed by a relatively slow electron capture process. During the second half period, the light is cut off. The conductance immediately dips down by $\sigma_{ph}$ due to the fast surface recombination, and then slowly decays by emitting electrons from the surface trap states. Analytically, one period of the red curve $\sigma(\omega t)$ in Fig.2b can be described by Eq. (1) below.

$$\sigma(\omega t) = \begin{cases} \sigma_T[1 - \exp(-t/\tau_c - T/(2\tau_c) + u/\tau_c)] + \sigma_{ph} + \sigma_d & -T/2 \leq t < 0 \\ \sigma_T \exp(-t/\tau_e + v/\tau_e) + \sigma_d & 0 \leq t < T/2 \end{cases} \quad \text{... Eq.(1)}$$

where $\tau_c$ is the capture time constant, $\tau_e$ the emission time constant, and the constants $u$ and $v$ (see the supplementary information for their expression) are set to ensure that the above equation is continuous. $\sigma_T$ is contributed by $\Delta p_1$ which is the excess holes left in the valence band after the corresponding electrons captured by trap states. $\sigma_{ph}$ is contributed by $\Delta n$ and $\Delta p_2$ in equilibrium. $\sigma_d$ is the nanowire conductance in dark and $\omega=2\pi/T$ with T as the chopping period.

According to advanced mathematics, any periodic function can be decomposed into an infinite summation of sine and cosine functions in the following form, so can eq.(1):

$$\sigma(\omega t) = \frac{a_0}{2} + \sum_{n=1}^{\infty}[a_n\cos(n\omega t) + b_n\sin(n\omega t)] \quad \text{... eq.(2)}$$

When the signal of above equation is fed into the lock-in amplifier, the amplifier multiplies the equation by $2\sin(\omega t)$ where $\omega=2\pi f$ and f is the chopping frequency (=1/T). All the terms are still in a waveform after the multiplication, except $b_n \sin(n\omega t)$ with n=1 which contains a dc component $b_1$. The internal filters inside the lock-in amplifier will filter out all the wave terms, allowing for only $b_1$, the amplitude of the fundamental term $\sin(\omega t)$, to be picked up. But the lock-in amplifier is designed to display the root mean square (RMS) instead of the amplitude of this term, meaning that the reading on the amplifier is $|b_1|/\sqrt{2}$. After transforming eq.(1) into the form of eq.(2), we find that $b_1$ can be expressed in the following form (see Section 1 of the supplementary materials):

$$b_1 = \frac{1}{\pi}\sigma_T\left[f\left(\frac{T}{2\tau_c}, \frac{T}{2\tau_e}\right) + f\left(\frac{T}{2\tau_e}, \frac{T}{2\tau_c}\right)\right] - \frac{2(\sigma_{ph}+\sigma_T)}{\pi} \quad \text{... eq. (3)}$$

where $f(a,b) = \dfrac{[1-e^a][1+e^b]}{[1-e^{a+b}][\left(\frac{b}{\pi}\right)^2+1]}$

It is widely known that, for trap states in silicon(8), the electron capture process is significantly faster than the electron emission process. In this case, we assume $\tau_c \to 0$, leading to

$$b_1 = \frac{\frac{\sigma_T}{\pi}\left[1+\exp\left(-\frac{T}{2\tau_e}\right)\right]}{\left(\frac{T}{2\pi\tau_e}\right)^2+1} - \frac{2(\sigma_{ph}+\sigma_T)}{\pi} \quad \ldots \text{ eq. (4)}$$

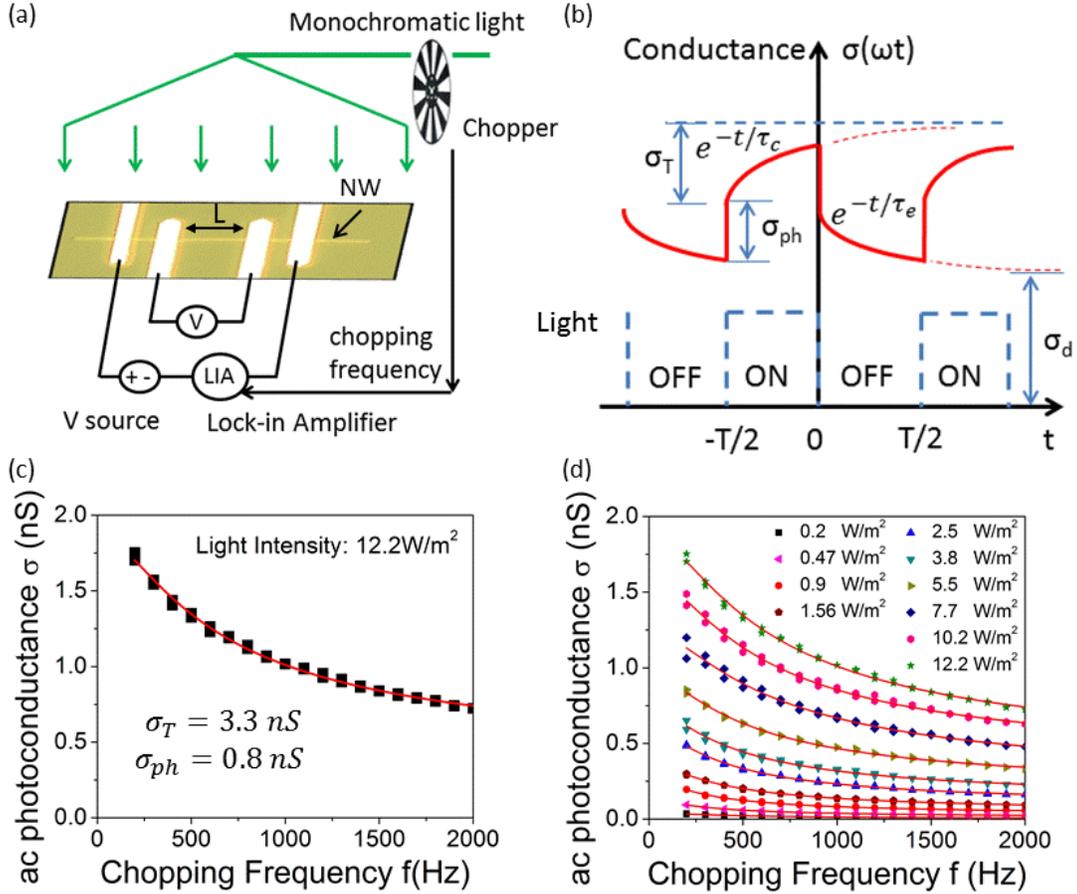

Figure 2. (a) Diagram of experimental setup. (b) Transit signal input to the lock-in amplifier. (c) Nanowire ac photocondance as a function of chopping frequency under illumination of 12.2 W/m². (d) Frequency dependent photoconductance under illumination of different light intensity.

By fitting $|b_1|/\sqrt{2}$ to the experimental data in Fig.2c, we find σ$_{ph}$ = 0.8 nS and σ$_T$ = 3.3 nS. The photoconductance σ$_{ph}$ comes from both the photoexcited electrons (Δn in Fig.1b) and their counterpart holes (Δp$_2$). But the trap-induced photoconductance σ$_T$ is only contributed by part of the total excess holes (Δp$_1$) in the valence band, because their corresponding photoexcited electrons are captured by the surface trap states. This fact allows us to conclude that, under illumination of a light intensity 12.2 W/m², the electron quasi-Fermi level $E_F^n$ of the nanowire shifts from the original location 0.47eV below to 0.23eV above the mid-bandgap energy level $E_i$, and that accordingly, a concentration $3.3\times10^{15}$ cm$^{-3}$ of electrons is captured by the trap states within the range of $E_F^n$ shift. The conclusion is based on the fact that the hole mobility μ$_p$ is estimated to be ~30 cm$^2$/Vs which is close to the hole mobilities of silicon nanowires found in literature (*14, 15*). The electron mobility μ$_n$ is assumed to be 3 times of this value.

To find the energy density of trap states, we sweep the quasi-Fermi energy across the bandgap by tuning the illumination intensity. The ac photoconductance as a function of the chopping frequency at different light intensity is illustrated in Fig.2d. All these curves are fitted by $|b_1|/\sqrt{2}$ of eq.(4) without any visible deviation, indicating that other effects such as local heating by illumination are negligibly small. From the fittings, we find a set of σ$_{ph}$ and σ$_T$, and plot them in Fig.3a and b, respectively. σ$_{ph}$ is linear with the light intensity, which shows that small injection condition is always maintained in the experiments. That is, Δn and Δp are both negligibly small compared to the majority holes, and therefore $E_F^p$ remains almost the same as $E_F$. Based on this, we conclude that only electrons are involved in the capture and emission processes of the surface trap states. The linearity of σ$_{ph}$ results in a logarithm dependence of $E_F^n$ on the illumination intensity, as shown in Fig.3a. It is evident that a small injection of illumination will shift $E_F^n$ over a wide range in the bandgap. If the density of trap states in this range is high, then a large number of carriers will potentially be trapped to induce a high photocurrent gain. This could be achieved by properly engineering the device surfaces.

Fig.3b shows the trap-induced photoconductance σ$_T$, which is a little nonlinear with the light intensity but always remains approximately 4 times of σ$_{ph}$ for the whole range. This photoconductance gain is mainly due to the fact that the trap states on the nanowire surfaces capture a large number of electrons. The same number of holes is left in the valence band to contribute to this amplified photoconductance. The capture of electrons on the nanowire surfaces will inevitably modulate the energy band of the silicon nanowire, introducing an additional change in photoconductance. Clearly, we cannot exclude the possibility that this "gating effect" (*16*) also plays a role in our device. The nonlinearity of σ$_T$ could be due to this effect. The "gating effect" induced by surface trap states has two properties that need to be pointed out. First, in time domain its modulation of photoconductance is similar to that of the capture-emission process by the trap states (Fig.2b). Second, how strong the "gating effect" is depends on the density of surface trap states, the initial net charge on the device surface and the size of the device (surface-to-volume ratio)(*16*), all of which are part of device surface properties. These facts make us believe that it is more appropriate to incorporate "the gating effect" as part of the capture-emission mechanism stated at the beginning of this Letter. By doing so, we can establish a unified model and quantitatively identify the nominal density of surface trap states that reflects all the properties of the device surfaces. The photoconductance gain induced by the "gating effect" is equivalent to an amplification of the density of surface trap states.

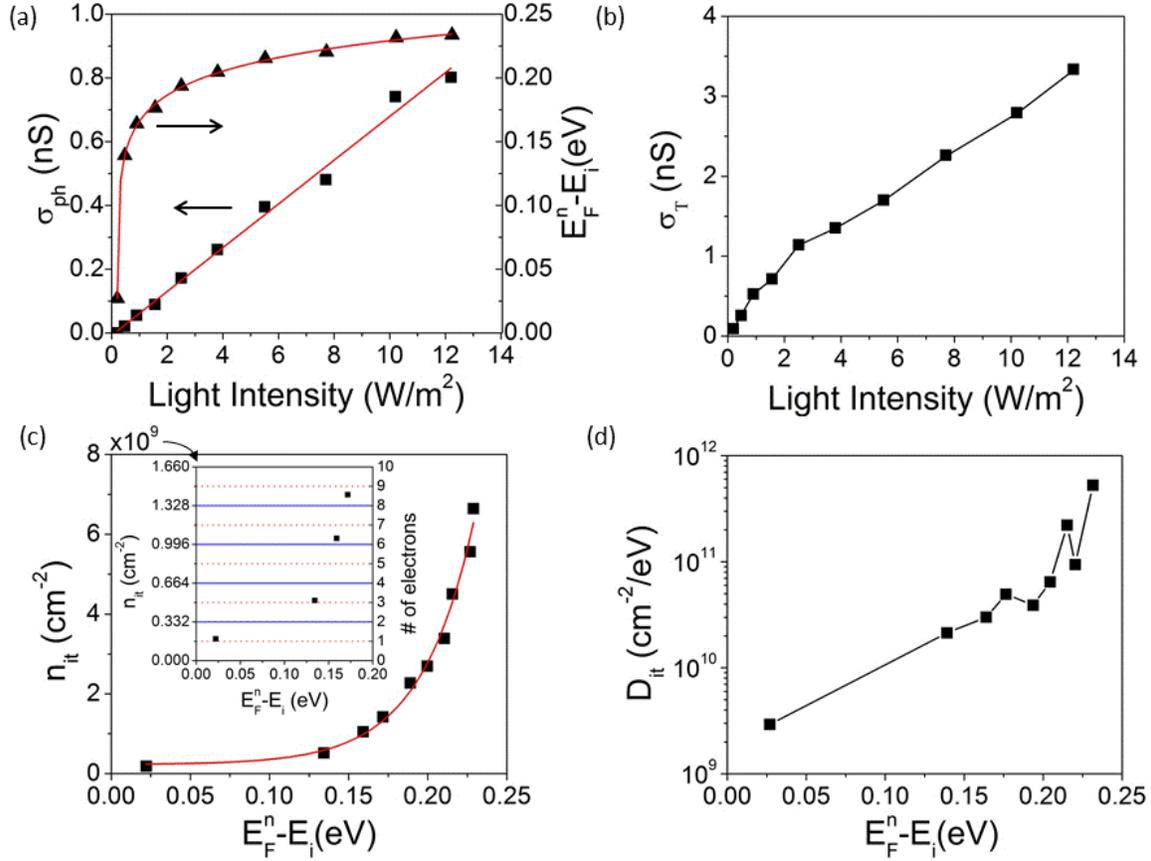

Figure 3. (a) Photoconductance σ$_{ph}$ and quasi-Fermi energy of electrons vs illumination intensity. (b) Trap-states-induced photoconductance σ$_T$ as a function of illumination intensity. (c) Surface concentration of trapped electrons as quasi Fermi level of electrons moves up. Inset: a close-up of the first four data points. (d) Calculated density of surface trap states of our nanowire device.

From Fig.3b, we calculate the number of excess holes (Δp$_1$) that contribute to the trap-induced photoconductance σ$_T$. The same number of electrons is captured by the surface trap states of the silicon nanowire. We plot the surface concentration of the trapped electrons as a function of $E_F^n$ shown in Fig.3c. The concentration follows an exponential curve. Amazingly, the first data point ($E_F^n - E_i = 22 meV$) has a concentration almost equal to the unit concentration (1.66×10$^8$ cm$^{-2}$) when only one electron is trapped on our nanowire surfaces (2.4μm long and 80nm in diameter), as shown in the inset. This indicates that there is one electron captured by the surface trap states. Accordingly, one counterpart hole is left in the valence to contribute to the photoconductance σ$_T$ in Fig.3b (see Section 2 of the supplementary materials). The electron concentration jumps up to only 3 times of this unit concentration (3 electrons captured) after $E_F^n$ increases by more than 110 meV (>4kT, k the Boltzmann constant and T=300K) to the second data point. This indicates that the captured electrons follow the Fermi-Dirac distribution instead of Boltzmann distribution. So it is most likely that there is one trap state located below $E_F^n$ of the first data point, and that two trap states are located within the gap between the first and second data point. A gap of 110 meV is wide enough to allow these two trap states to be spaced far apart so that they can be filled with electrons one by one. This is the reason why we observed the quantization in the charge trapping. At high energy levels, the trap states are located more closely next to each other. The quantization of charge trapping becomes increasingly less likely (see the inset).

We calculate the density of surface trap states by simply differentiating the surface concentration of trapped electrons over the quasi Fermi energy, as shown in Fig.4d. The trend is the same with what is obtained by the C-V characteristics of nanowire MIS capacitors(*7*), but our obtained density is 2 orders of magnitude more precise. At deep levels, the density is accurate since the filling of trap states is quantized. At higher energy levels where the energy gaps between trap states are narrower, the density is somewhat overestimated since the electrons have high probabilities to be captured by the states above $E_F^n$. An accurate solution is always difficult to find since the related equation is ill-defined (see Section 3 of the complementary materials for details).

In conclusion, we have demonstrated a simple but powerful method to accurately measure the density of surface trap states at single nanowire level. At room temperature, we observed that the charge trapping by the trap states is quantized, indicating that the states are filled with electrons one by one. This method allows us to conveniently probe the density of trap states in ultra-scaled nano/quantum devices at very high precision, which is not possible by traditional capacitive techniques. It opens up opportunities to develop high-performance nanodevices by engineering the device surfaces.

**Methods**

The silicon nanowires were synthesized by the typical gold-catalyzed vapor-liquid-solid (VLS) process in a mixture of $H_2$-balanced $SiH_4$ and $BH_3$. To make contacts to single nanowires, we first prepared a nanowire suspension by sonicating the as-grown nanowire samples in isopropanol (IPA) followed by a proper cleaning process(*17*). A drop of the nanowire suspension was then applied onto a quartz substrate with prefabricated alignment markers. The existence of these markers enables us to locate individual nanowires under optical microscope. Ohmic contacts were made to the individual nanowires by multiple microelectrodes (200nm Pd with 3nm Cr as the adhesion layer) formed in the processes of photolithography and metallization. After proper wire-bonding, we placed the dc biased nanowire device under illumination of a monochromatic light beam (λ ≈ 500 nm) that is modulated ON/OFF periodically by a mechanical chopper. The illumination intensity on the nanowire device is difficult to accurately estimate, and are subject to a variation of a few times in the related figures. But this variation does not affect the conclusion we reached in this Letter. A Labview script was used to tune the modulation frequency from 200 Hz to 2000 Hz, and the ac photocurrent was picked up by a lock-in amplifier. We performed four probe measurements on the nanowire device to exclude the contact resistance.

**Acknowledgements**

We thank Prof. Ali Javey for providing us silicon nanowires, and Prof. Abdelmadjid Mesli and Dr. Siew Li Tan for useful discussions. The work reported here is supported by the national "1000 Young Scholars" program of the Chinese central government, the National Science Foundation of China (61376001) and the "Pujiang Talent Program" of the Shanghai municipal government.

**Completing financial interests**

The author declares no completing financial interests.